# A micro-optical module for multi-wavelength addressing of trapped ions

# A micro-optical module for multi-wavelength addressing of trapped ions

Matthew L. Day,[1,2,3,†] Kaushal Choonee,[2] Zachary Chaboyer,[4] Simon Gross,[4] Michael J. Withford,[4] Alastair G. Sinclair,[2] and Graham D. Marshall[1]

1 Quantum Engineering Technology Labs, University of Bristol, BS8 1TL, UK
2 National Physical Laboratory, Teddington, TW11 0LW, UK
3 Quantum Engineering Centre for Doctoral Training, University of Bristol, BS8 1TL, UK
4 MQ Photonics Research Centre, Department of Physics and Astronomy, Macquarie University, New South Wales 2109, Australia

Email: alastair.sinclair@npl.co.uk

**Abstract**

The control of large-scale quantum information processors based on arrays of trapped ions requires a means to route and focus multiple laser beams to each of many trapping sites in parallel. Here, we combine arrays of fibres, 3D laser-written waveguides and diffractive microlenses to demonstrate the principle of a micro-optic interconnect suited to this task. The module is intended for use with an ion microtrap of 3D electrode geometry. It guides ten independent laser beams with unique trajectories to illuminate a pair of spatially separated target points. Three blue and two infrared beams converge to overlap precisely at each desired position. Typical relative crosstalk intensities in the blue are $3.6\times10^{-3}$ and the average insertion loss across all channels is 8 dB. The module occupies $\sim10^4$ times less volume than a conventional bulk-optic equivalent and is suited to different ion species.

## 1. Introduction

Exquisite manipulation of light is essential for the advancement of technologies reliant on the quantum properties of atomic particles [1]. The full potential of these physical systems will be realised by using scalable components; micro- and nano-fabricated devices are the ideal means for this. Examples of such devices for cold atomic systems include 2D diffraction gratings as atom chips [2], a waveguide chip to realise an array of atom-photon junctions [3], micro-ring resonators as a route towards a scalable atom-light interface [4], and a nanophotonic waveguide circuit for a configurable optical tweezer array [5]. In the field of quantum photonics, adopting a modular component approach permits the use of the most appropriate materials for the desired functions [6] as well as providing the flexibility to construct adaptable systems [7]. In atomic-based technologies, micro- and nano-photonic modules have clearly advanced the development of miniaturised atomic instruments such as stabilised lasers [8] and clocks [9].

---

[†] Present address: Institute for Quantum Computing, University of Waterloo, N2L 3G1, Canada

The viability of trapped-ion arrays for quantum information processing [10-13] has stimulated much of the development of microfabricated trap devices [14] for scaling up to large numbers of ion qubits. The quantum CCD architecture is one promising way of scaling up trapped ion quantum information processors to useful sizes with ions distributed and shuttled amongst many different trapping zones [15, 16]. Laser beams are a prerequisite to initialise and detect the ions' quantum states and are often used to implement quantum operations; yet very few efforts towards scalable optical systems, capable of irradiating multiple trapping sites in parallel in a microtrap device, have been reported [17-19]. For full control of ions at a single trapping site, multiple laser beams are required to irradiate that site. Tuned to different wavelengths, these lasers are used for cooling and repumping as well as coherent control of the ion's electronic and motional energy levels. Often beams at a single wavelength with different $k$-vectors are required; e.g. a cooling trio for sensing micromotion in 3D [20], or a pair for qubit control. Furthermore, these control techniques require exact $k$-vectors and polarisation states of specific beams with respect to the quantisation and trap axes. A system of laser beams focused by bulk optics is not a scalable means to optically address an array of trapping zones. Speaking generally, a micro-optical system created using microfabrication techniques offers a promising route to a practical solution.

The most sophisticated demonstrations to date used planar waveguides and focusing grating couplers integrated within the substrate of a 2D surface-electrode microtrap. This principle was shown first for 674 nm ($^{88}$Sr$^+$ optical qubit) [18] and is now extended to all wavelengths for $^{88}$Sr$^+$ [17] as well as to 729 nm ($^{40}$Ca$^+$ optical qubit ) [21]. These microfabricated 2D surface-electrode traps with integrated photonic elements offer a viable route to scaling the number of optically addressed trapping zones in these devices. The key advantage of fabricating the electrodes, waveguides and focusing couplers in a monolithic structure is the intrinsic referencing of the ion and optical beams with respect to each other. This enables high stability optical addressing as all vibrations are common mode, however the placement of grating coupler elements on the trap surface may constrain electrode layouts.

In contrast and comparison to 2D surface-electrode devices, microtraps with a 3D electrode geometry [22] benefit from greater trap depth, higher harmonicity potentials and lower heating rates. These 3D devices would also benefit from the use of integrated optical techniques to achieve a scalable optical interconnect. To include photonic elements in the 3D micro-structure itself increases considerably the technical complexity of the device and the associated fabrication risks, as well as restricting achievable beam geometries. Combining an optical interconnect module to a 3D microtrap in a hybrid structure permits these components to be optimised separately, so that the most suitable materials and architecture for optimum performance can be chosen for each component. It also permits the species-specific wavelength requirements to be confined to the optical module only, whereas the same microtrap device structure can be applied to all species. This modular assembly will require the two devices to be aligned and bonded using photonics or MEMS packaging techniques; this is a challenge not present in the 2D surface electrode devices mentioned above.

This paper presents the design, development and characterisation of an efficient micro-optical module, the principle of which is scalable to multiple trapping zones of an ion microtrap array of 3D micro-structured form. We combined femtosecond-laser direct written (FLDW) waveguides and transmissive diffractive optics, to create an optical module that delivers multiple focused and overlapped laser beams, at blue and near infrared wavelengths, with $k$-vectors suited to microtrap operation. FLDW waveguides are chosen such that waveguides for supporting optimal single-modes of blue to NIR wavelengths can be fabricated in a single step, within a monolithic piece of glass. We fully characterise the fabricated device and assess its suitability for interfacing with ions in a microtrap according to the optical efficiency, degree of laser beam overlap, and the degree of optical crosstalk between adjacent beam overlap sites.

## 2. Micro-optic module

### 2.1 Module concept and design

The module architecture is illustrated in Figure 1(a). A 1D fibre v-groove array (VGA) contains various input optical fibres suitable for single-mode operation at the wavelengths required. A waveguide array remaps the VGA fibres' mode positions (see Figure 1(b)) to those of a 2D array of diffractive microlenses (DMLs). The DML array is prepared on a separate substrate optimised for the manufacture of the DMLs. The output modes from the waveguide array propagate unguided in the DML substrate and expand before focusing, thus increasing the numerical aperture (NA) of the focusing geometry. The individual elements are intended to be in direct contact with each other, as well as being bonded together to ensure long-term stability. The module was developed specifically for an ion microtrap with a 3D electrode geometry; all beams are intended to propagate through the microtrap aperture (see cross section in Figure 1(c)). A micromachined spacer will be required to ensure a 2 mm working distance from the ions and facilitate bonding the module to the microtrap chip. The positions of the DMLs are chosen to generate the required laser beam $k$-vectors and meet the geometrical requirements of the trapping zones of the 3D microstructure, such as in the device shown in Figure 1(d). With this arrangement, the module will be contained within UHV. The technical challenges to realise operation of the module with a microtrap are discussed in section 4.

DMLs were chosen because a) they provide a means to fabricate lenses in vacuum-compatible materials using lithographic techniques, b) they have the capacity to redirect the light in an off-axis configuration, and c) access to in-house EBL and etching tools permitted rapid iterative optimisation of the lens fabrication process. Refractive microlenses are also a possibility; state-of-the-art fabrication of thick refractive microlenses (up to 400 µm) in UHV compatible materials is performed using direct laser micromaching [23]. This process is commercially available but iterative optimisation would be slower than using in-house facilities.

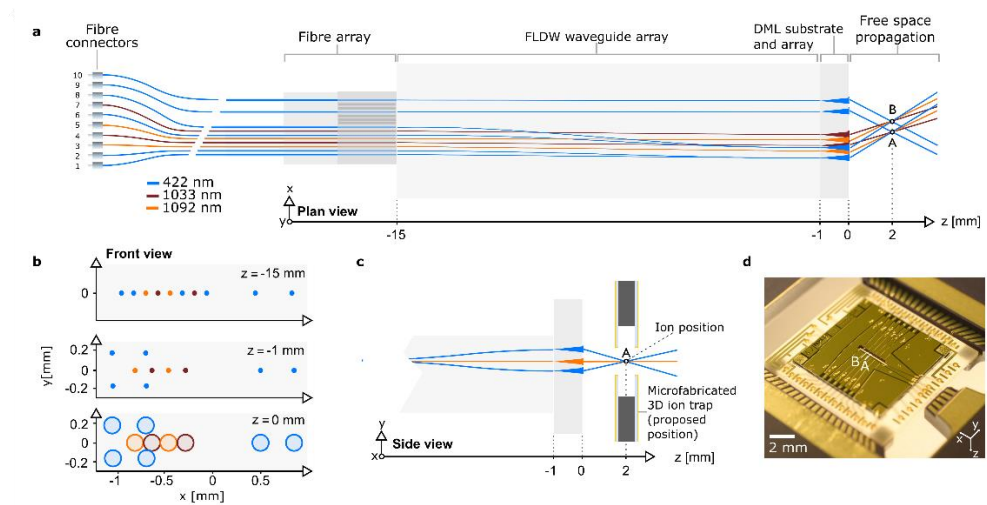

Fig 1. Overview of the micro-optical module. **a.** Plan view of the module concept and demonstrator design. Light is injected from the left-hand side into the fibre connectors which route the light to a FLDW waveguide array. The FLDW waveguides remap the fibre core positions to the ($x,y$) positions of a lens array. The lens array directs and concentrates the light at the two addressed sites, A and B, at $x = \pm 175$ µm, $y = 0$, $z = 2000$ µm, where $z = 0$ µm is the DML array plane. **b.** Layout of beam locations at the input ($z = -15$ mm) and output ($z = -1$ mm) of the waveguide array, and at the DML array plane ($z = 0$ mm), demonstrating the remapping of the fibre array outputs to lens positions (waveguide diameters have been exaggerated for clarity) **c.** Side view of the front section of the module demonstrating the design orientation of the module with respect to an ion microtrap. Use of the module with a microtrap will require further development (see Section 4). **d.** The ion microtrap array with matching site labels.

Our efforts concentrated on demonstrating operation of the module at the wavelength extremes relevant to the control of $^{88}$Sr$^+$. This ion species requires 422 nm (Doppler cooling and electronic ground state initialisation), 674 nm (coherent control on the optical qubit transition), 1033 nm (quenching the metastable excited state of the qubit transition) and 1092 nm (repumping during Doppler cooling) [24]. Additionally, 461 nm and 405 nm are used for photoionisation of Sr atoms [25]. Table 1 presents the module design parameters for the two beam sets addressing two microtrap array sites (A & B, separated by 350 µm). Each beam set contains one quenching and one repumping beam, as well as three non-coplanar Doppler cooling beams (a geometry useful for sensing micromotion in 3D using the RF-correlation technique [20, 26]). Although developed for $^{88}$Sr$^+$ ions in an existing microtrap [22], the principle of this work is applicable to other devices and atomic species (see Discussion).

Table 1. Summary of the beams in the optical assembly. The beam name is shortened to DC—Doppler cooling, Q—Quench and R—Repump with the site location in parentheses. There are three separate DC beams (numbered 1 to 3) for each site. The waveguide number, VGA port, wavelength, designed unit *k*-vector are shown. The waveguides and VGA ports are numbered from left to right in the x-axis with the VGA port number corresponding to the position in a 16-groove array with 127 µm pitch. The measured $1/e^2$ beam diameters at the intersection points are included on the right-hand side (see Section 3.2).

| Beam name | Waveguide | VGA port | λ (nm) | Designed unit *k*-vector | Measured beam diameter, $2w_{x,y}$ (µm) |
|---|---|---|---|---|---|
| DC1(A) | 1 | 1 | 422 | $0.40\hat{x} + 0.07\hat{y} + 0.92\hat{z}$ | 9.8, 6.8 |
| DC3(A) | 2 | 2 | 422 | $0.40\hat{x} - 0.07\hat{y} + 0.92\hat{z}$ | 9.4, 6.8 |
| R(A) | 3 | 3 | 1092 | $0.30\hat{x} + 0\hat{y} + 0.95\hat{z}$ | 22.1, 22.6 |
| Q(A) | 4 | 4 | 1033 | $0.22\hat{x} + 0\hat{y} + 0.98\hat{z}$ | 19.0, 19.9 |
| R(B) | 5 | 5 | 1092 | $0.30\hat{x} + 0\hat{y} + 0.95\hat{z}$ | 22.2, 23.3 |
| DC1(B) | 6 | 6 | 422 | $0.40\hat{x} + 0.07\hat{y} + 0.92\hat{z}$ | 9.7, 6.3 |
| Q(B) | 7 | 7 | 1033 | $0.22\hat{x} + 0\hat{y} + 0.98\hat{z}$ | 19.7, 20.4 |
| DC3(B) | 8 | 8 | 422 | $0.40\hat{x} - 0.07\hat{y} + 0.92\hat{z}$ | 8.9, 6.5 |
| DC2(A) | 9 | 12 | 422 | $-0.32\hat{x} + 0\hat{y} + 0.95\hat{z}$ | 7.2, 6.9 |
| DC2(B) | 10 | 15 | 422 | $-0.32\hat{x} + 0\hat{y} + 0.95\hat{z}$ | 7.4, 6.8 |

*2.2 Laser-written waveguide chip*

Ten connectorised fibres for single-mode operation at the channel design wavelength terminate in a VGA (127 µm pitch, *OzOptics*) for coupling to waveguides in the chip; S405-XP and 1060XP fibres (both *Nufern*) were used for blue and NIR wavelengths respectively. At short wavelengths, the specified maximum core position deviation in the VGA (±1.5 µm, which includes core-cladding offset and fibre position deviations within the VGA) can cause a significant coupling loss. To minimise this loss, the measured fibre core positions (±0.2 µm) supplied by the vendor are used as the input locations of the waveguides; 3D positioning of the FLDWs makes this possible. We find that waveguide position compensation, although not strictly necessary, improves overall transmission efficiency by 10-30% (see supplementary material). The waveguide chip maps input modes from the 1D VGA to the positions of microlenses arranged in a 2D array. The design of 3D waveguide trajectories and assignment of fibre ports to a lens was performed using the Kuhn-Munkres algorithm to globally maximise the bend radii of the waveguides' cosine curve trajectories. A curved remapping region of 12 mm length inside the chip ensured all waveguide bends (radii > 40 mm) were low-loss [27]. The exact overall chip length is unimportant and propagation losses were low (see Section 3.1) so short straight lead in and out waveguide sections were included before and after the remapping bends to ease tolerances when preparing the chip facets.

The FLDW waveguides are written by focusing a Ti:Sapphire laser ($\lambda$ = 800 nm, repetition rate = 5.1 MHz, pulse duration < 50 fs) inside a 30 mm × 14 mm × 1.1 mm glass substrate (Eagle2000) using an oil-immersion objective lens (100×, 1.25 NA). In the laser focus, non-linear absorption causes a permanent and positive refractive index change [28-30], so translating the substrate creates a waveguide with a 3D trajectory. Fabrication process control was achieved by operating in the cumulative heating regime, where the laser repetition period is shorter than the thermal relaxation time of the glass and waveguide diameters increase approximately linearly with laser fluence [31-33]. After exposure, thermal annealing relaxed stress around the densified waveguide region, forming approximately Gaussian refractive index profiles. The FLDW waveguide cores have an approximately circular symmetry resulting in low birefringence [27, 34]. We ensured minimised birefringence by applying a more stringent thermal annealing ramp than in [34], which reported a birefringence of $1.2 \times 10^{-6}$. Our mode radius measurements showed the same ellipticity as in [34], so we used this value to estimate a polarisation cross-talk of -12 dB for our waveguides. The blue (NIR) 3D waveguides were written with ~40 nJ (~70 nJ) pulses and a 1000 mm/min (500 mm/min) feedrate. Rapid writing (~1 guide /s) enables large arrays and multiple samples to be made in short timescales. Empirically determined pulse energy adjustments ensured constant diameter waveguides at different depths (60 µm – 400 µm). After the chip facets were lapped and polished, the waveguide arrays were tested before permanently attaching to VGAs using UV-curing optical adhesive. To the best of our knowledge, the blue FLDW waveguides developed here are at the shortest operating wavelength for curved, 3D-trajectory waveguides.

### *2.3 Diffractive microlens array*

DMLs were designed using scalar diffraction theory [35], with the accuracy of the design process confirmed using vector simulations in COMSOL Multiphysics®. More advanced design procedures could be employed to increase diffraction efficiency of the microlenses [36] which would lower the optical crosstalk. The choice of *k*-vectors for the two triplets of non-coplanar 422 nm beams (see Table 1) determined the locations of the associated DMLs. Following this, the *k*-vectors for the two NIR beam pairs were chosen so that the associated DMLs did not overlap with any of those for the 422 nm beams. DML diameters (170 µm) were chosen to achieve a NA ~0.13, thus ensuring efficient coupling and minimal scatter of light emerging from the waveguide modes. DML designs were generated with the waveguide output position as the object point, and with the desired microtrap zone to be addressed as the image point. The designed phase profiles were then converted into $TiO_2$ surface-relief profiles for the fabrication procedure.

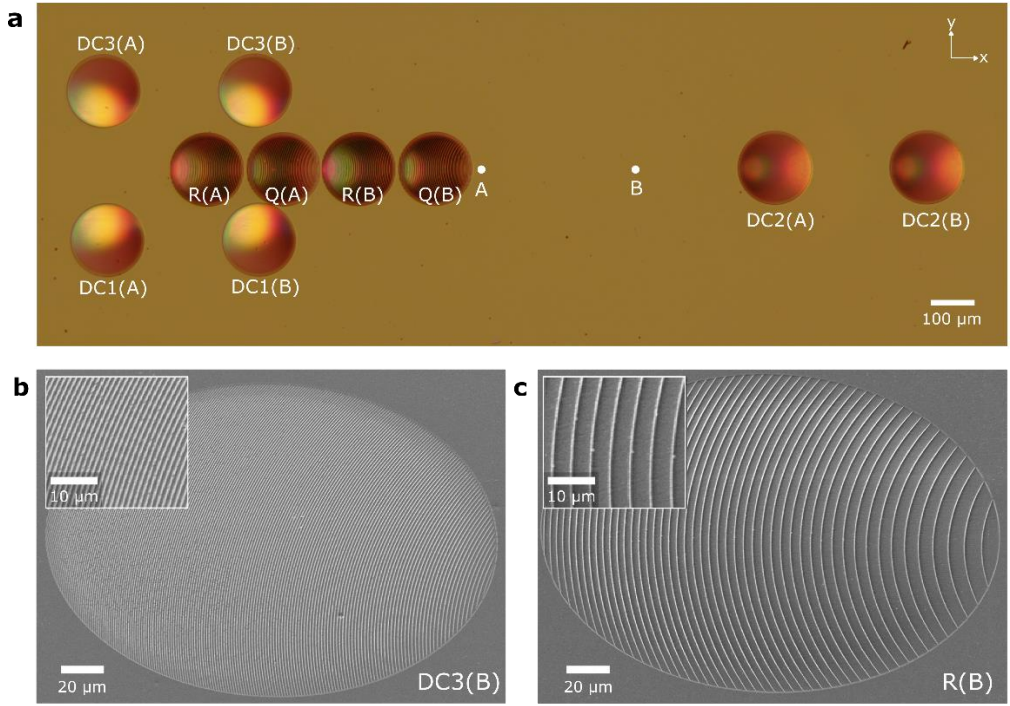

Fig 2. Images of the fabricated DML array in TiO$_2$ thin film labelled with their corresponding beams as defined in Table 1. **a.** An optical microscope image with the intersection points A and B projected onto the DML plane. **b,c.** SEM images of two DMLs which have been imaged at 45 degrees. The insets are magnified views of the central ~30 µm of the DMLs. (Ring-like structures in these images are aliasing artefacts and are not present in the DMLs themselves.)

Fused silica substrates (($1030 \pm 20$) µm thick) with an amorphous TiO$_2$ film (996 nm thick) were coated with a PMMA resist before metallisation (13 nm Al film) to prevent charging during electron exposure. Electron beam lithography (EBL) exposed greyscale dose maps (25 nm pixels) which were compensated for the non-linear response of PMMA to produce the desired surface-relief structures after Al removal and subsequent resist development (propan2ol:H$_2$O). Plasma etching (CF$_4$/Ar) transferred the PMMA surface-reliefs into the TiO$_2$ film. Distortion effects due to reduced transport into narrow grating lines during the etch process were compensated for in the electron exposure. The design of the ten-DML array assumed the upper bound of the substrate thickness to ensure that the array could be positioned optimally in the z direction. Figure 2 shows the fabricated array used in the micro-optical assembly.

### 3. Characterisation of module performance

Optical testing of sub-components and the complete micro-optical module was performed using lasers at $\lambda = 422$ nm and 1033 nm and 1092 nm, delivered by single-mode fibre, with optical powers of ~1 mW. These powers were chosen to avoid degradation of the UV epoxy, especially from blue wavelength light. The DML array and the bonded fibre-waveguide array were mounted on 3-axis and 6-axis flexure stages respectively, to optimise alignment. To enable maximum flexibility of combining available samples for these and future experiments, the DML and waveguide chips were not bonded, although this could be done straightforwardly and would be in a module interfaced with a microtrap.

## 3.1 Optical efficiency

The insertion loss of the full module, defined as the reduction in transmission from the fibre array to the ion position, was measured to be 7.8 dB when averaged over all channels. This loss can be directly compared to losses of 16 dB to 25 dB in a planar waveguide system at the same wavelengths [17]. To characterise the sources of loss, we measured the transmission at the output of each component identified in Figure 1(a). The average insertion loss for each wavelength in each component is presented in Table 2. Sources of loss in the waveguide chip are twofold: fibre-to-waveguide coupling, and propagation. The insertion losses of the waveguide array are lower bounded by the overlaps ($S$) between the fibre and waveguide modes. At 422 nm, 1033 nm and 1092 nm the overlaps were measured to be $S_{422} = 98\%$, $S_{1033} = 80\%$ and $S_{1092} = 71\%$ respectively (see supplementary material).

Table 2. Average insertion loss in each component as defined in Figure 1(a) for each operational wavelength. The combined loss of the waveguide chip and the DMLs represent losses intrinsic to the device. The excess loss at the fibre connector loss was due to a slight mismatch between the VGA fibres and the patch fibre cables used to connect to the test laser sources. The most significant loss in the system occurs within the DML array, both through reflection and reduced diffraction efficiencies. These two loss mechanisms can be improved to increase the overall transmission of the device.

| Component | Average insertion loss (dB) | | |
|---|---|---|---|
| | $\lambda$=422 nm | $\lambda$=1033 nm | $\lambda$=1092 nm |
| Fibre connector | 5.4 | 3.1 | 3.7 |
| Waveguide array | 1.3 | 1.8 | 1.4 |
| DML reflection | 3.2 | 2.5 | 2.5 |
| DML diffraction efficiency | 3.1 | 3.1 | 4.4 |
| Total | 13.0 | 10.5 | 12.0 |
| Total (excluding fibre connector) | 7.6 | 7.4 | 8.3 |

To avoid destructively reducing the chip length, waveguide propagation loss was not measured. Nevertheless, the intrinsic material absorption is low at the operating wavelengths [37]. No correlation between the minimum bend radii of the waveguides and their insertion loss was observed, thus bend losses are not dominant. By assuming that all waveguide chip losses occur at the fibre-waveguide interface, we determine the upper bound alignment error between the fibre and waveguide core modes to be <700 nm at $\lambda$ = 422 nm, which demonstrates the excellent registration between the components. The waveguides for NIR wavelengths exhibit slightly higher insertion loss than for 422 nm, which is likely due to a combination of the decreased mode overlap and the modes propagating in the weakly guiding regime. Both these sources of loss could be improved by increasing the diameter of the waveguides. The parameters for NIR waveguides were developed using light at $\lambda$ = 976 nm (due to test laser availability at the fabrication facility). Extrapolating from these fabrication parameters enabled writing of guides for 1033 nm and 1092 nm. Improved performance could be achieved with the availability of lasers at the exact NIR wavelengths during fabrication development. The NIR waveguides in the device presented have bend radii > 250 mm, such that their bend loss is negligible. We expect all losses due to weak guiding to be radiative.

Losses in the DML array are due to diffraction inefficiency and reflection from the substrate-air interface. Due to the high refractive index of the $TiO_2$, the reflection of the incident radiation at the DML plane is strongly dependent on the post-fabrication film thickness. We measured the total reflection losses from the DML array substrate to be 3.2 dB and 2.5 dB for blue and NIR wavelengths respectively. Transmission could be maximised via improved process control of the $TiO_2$ film thickness. The DML diffraction efficiencies (equivalent insertion losses) were measured to be 49% (3.1 dB) for λ = 422 nm and λ = 1033 nm, and 36% (4.4 dB) for λ = 1092 nm. The DMLs for NIR wavelengths were under-etched due to insufficient thickness of EBL resist to realise the required ~1 µm surface-relief depths; a thicker resist, or increased etch selectivity, could improve the etch process and consequently the diffraction efficiency.

## *3.2 Laser beam overlap*

Beam profiles of the focusing laser beams, imaged in the intersection plane (i.e. interface 4 in Figure 1b), are shown in Figure 3. Both sets of five beams overlap at their corresponding designed intersection points which are separated by 350(1) µm, at a distance 1990(10) µm from the DML plane. At both points the 422 nm beams intersect within the measurement resolution (1 µm) except for one beam, which is offset by ≤ 2 µm. This offset corresponds to a relative intensity $I/I_0 = 0.89$ for that beam at the design intersection point, where $I_0$ is the peak intensity. The repeatable and accurate overlap of the 422 nm beams demonstrates the excellent agreement between the design model and the fabrication process used for these DMLs. The sequence of measured profiles of a 422 nm beam set as it propagates through intersection point A is shown in Figure 4, demonstrating the overlap of three beams at a single point in space. The 1033 nm and 1092 nm beams have slightly imperfect overlaps with the design intersection points and are systematically displaced at both sites by 7.5 µm and 5 µm respectively from the 422 nm beam locations. These small systematic errors result in corresponding relative intensities at the intersection points of $I/I_0 = 0.69$ and 0.93 respectively. This imperfection could be readily corrected (see section 4.2).

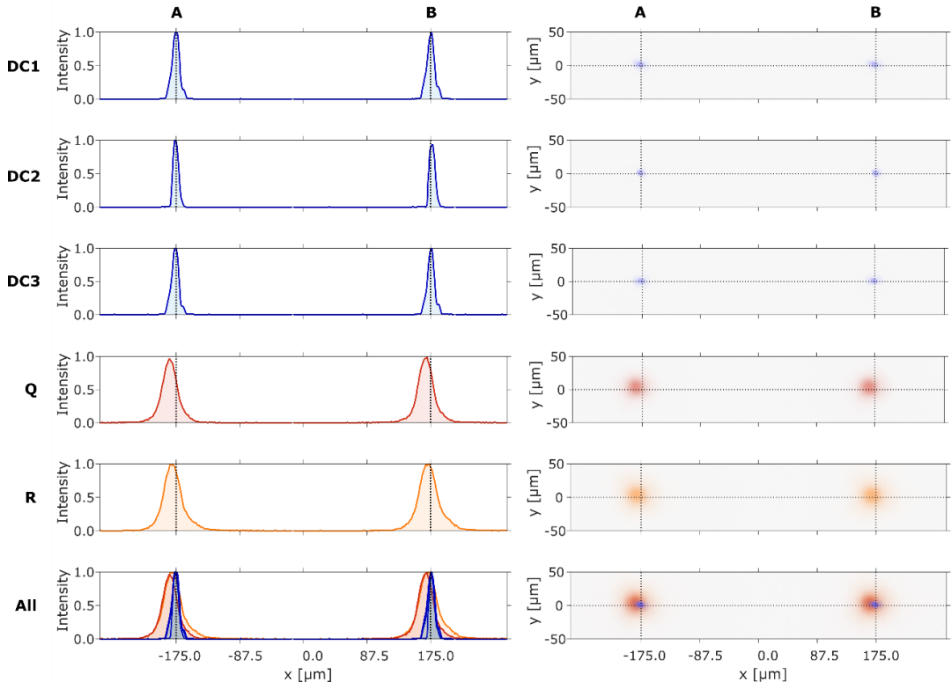

Fig 3. Cross sections of the beam profiles (left) and intensity maps in the intersection plane (right). The designed intersection sites A & B are shown at $x = -175$ µm and $x = +175$ µm respectively. The blue beams (DC1,2,3 at 422 nm) are in good agreement, while the NIR beams have relative intensity reductions at the designed intersection point of 31% (Q at 1033 nm) and 7% (R at 1092 nm).

The measured $1/e^2$ diameters of the beams (2*w*) in the intersection plane are given in Table 1. These dimensions are taken in the *x* and *y* axes to enable meaningful comparison of all the beams. The beam diameters are reproduced across the two sets of beams which demonstrates the consistency of the fabrication process used to create the waveguides and the DMLs. The measured diameters are approximately twice the size that would be expected from simulations of the ideal DML phase profiles. Simulations of DMLs with imperfect phase profiles indicate that beam diameter broadening can be explained by a reduction in the DML depth with pitch, such that a full 2π phase shift is not achieved away from the centre of the DML. The demonstrated beam profiles are suitable for few-ion addressing (i.e. 2 to 3 ions, not single ions) at each microtrap zone. This configuration is relevant to many architectures in trapped ion quantum technologies, for example as in the work of Pino *et al*. [16].

The measured profiles for all beams exhibit non-Gaussian features. For the 422 nm beams, this is likely due to comatic aberrations from the large (~ 20°) off-axis deflection, combined with intersection point not being exactly at the focal point of the diffractive lenses. The NIR beams have Lorentzian tails, consistent with measurements of the waveguide mode profiles (see supplementary material), which can be corrected by fabricating larger diameter waveguides.

### *3.3 Optical crosstalk*

When performing quantum state readout on one or more sites in an array, an important consideration is the intensity of crosstalk light at unaddressed sites. This light may emanate from waveguides, their interfaces, or higher diffraction orders of the DMLs. Light contamination at 422 nm is a critical aspect; this wavelength connects to a qubit level and ions at an unilluminated trapping site could scatter these photons during state readout elsewhere. A single scattering event is sufficient to destroy the coherence of the system [1]. Therefore by performing quantum state readout in one trapping site, crosstalk could cause decoherence of ions at other trapping sites. This is an important concern for measurement-based protocols, where state detection determines operations conditioned upon the measurement outcome [13, 38].

Beam profiles recorded using a high-dynamic range procedure (see supplementary material) are shown in Figure 5; these demonstrate relative crosstalk intensities $I_{\text{cross}}/I_0 < 5.4 \times 10^{-4}$ at the adjacent intersection points for all the 422 nm beams. To assess the applicability of the device to larger numbers of trapping zones crosstalk intensities were also measured in the region up to $\Delta x = \pm 1300$ µm from the illuminated intersection point. This is important because it includes the locations where un-diffracted light will occur. Indeed, in all measurements un-diffracted light was the source of the largest background intensity. In the realistic case when ions are stored at separated trapping sites, there will be one microtrap electrode zone between the two sites to provide an electric potential barrier. At these distances of ±700 μm from the illuminated site, crosstalk intensities of $< 5.5 \times 10^{-4}$ were achieved and, over the whole of the measured region, the highest crosstalk intensity was $< 3.6 \times 10^{-3}$ (with one exception of DC3(A) which exhibited $6.7 \times 10^{-3}$ in the worst instance; see supplementary material). The intensity profile of DC1(A), with *x* quantified in units of beam radius *w*, achieves a similar *I/I*$_0$ as the data presented by Mehta and Ram [39] out to *x* = 4*w*, where our noise floor limits the measurement. Sensitivity down to *I/I*$_0$ ~10$^{-5}$ is required for a more complete comparison.

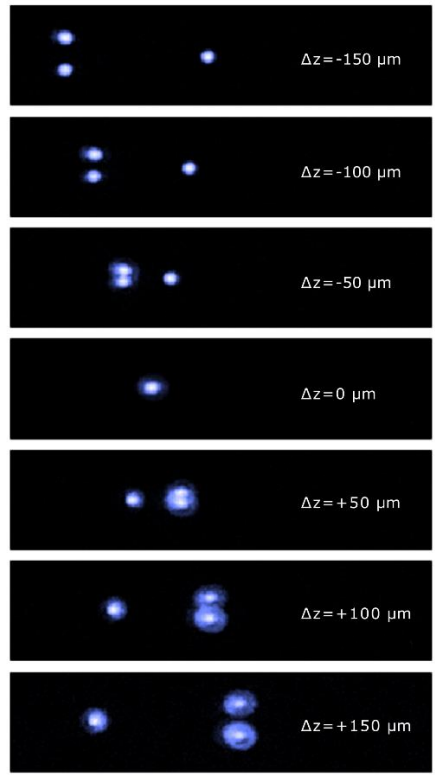

Fig 4. Sequential images of the three 422 nm beams propagating through intersection point A in 50 µm steps. Identical behavior is observed at intersection point B, which occurs in the same plane as intersection point A. Each image has a 340 μm × 80 μm field of view.

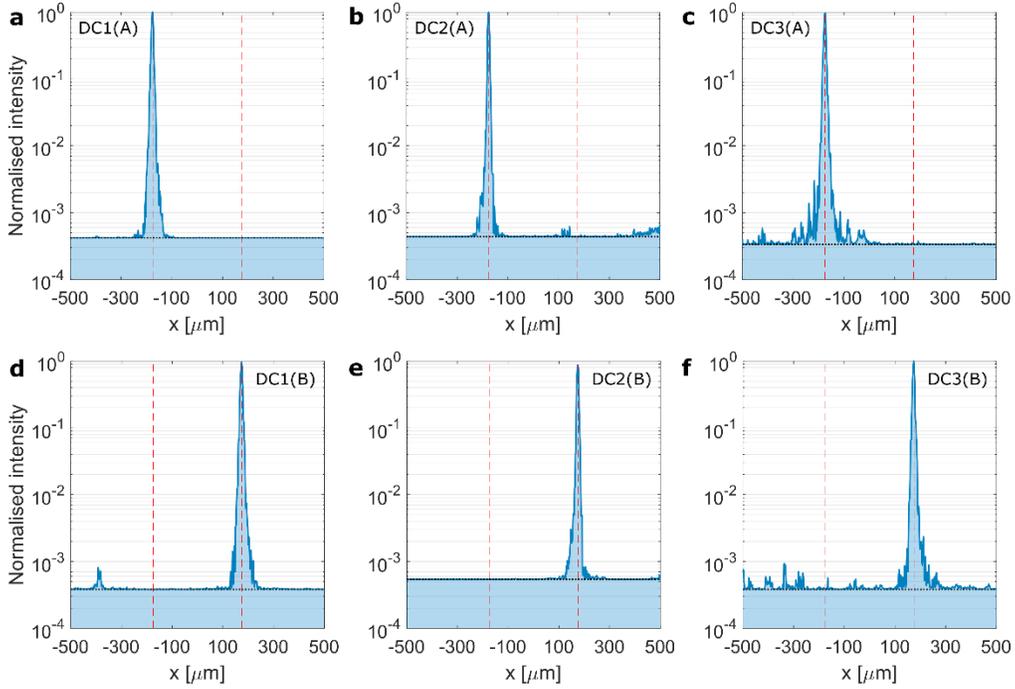

Fig 5. High dynamic range beam profiles of all 6 Doppler cooling beams, **a.-c.** DC1(A), DC2(A) and DC3(A) respectively, **d.-f.** DC1(B), DC2(B) and DC3(B) respectively. The red dashed lines indicate positions A ($x = -175$ µm) and B ($x = +175$ µm). The minimum normalised intensity detectable in the measurement scheme is represented by the black dotted line and was limited by the detector's dark current noise relative to the peak intensity measured for each beam profile. The relative noise floor for each lens varies due to slight variations in the peak intensity for each beam.

## 4. Discussion

### 4.1 Applicability to ion species

A primary requirement for the optical module is guidance of laser light at the typical wavelengths needed for laser cooling of trapped ions and coherent control of qubit states. The module presented here concentrated on spanning the wavelength range of transitions in $^{88}Sr^+$. Waveguides and lenses were demonstrated successfully at the extremes of this range. Waveguides and DMLs for 674 nm for the qubit transition in $^{88}Sr^+$ could be created in a future device of the same materials by interpolating the fabrication parameters established for the present wavelengths. Beyond $^{88}Sr^+$, the principles demonstrated in this work are applicable to other ion species. For example, $Ca^+$ has an analogous electronic fine structure with transitions at 397 nm, 729 nm, 854 nm and 866 nm (cool, qubit, quench, repump respectively) and is widely used for quantum information research [12, 40]. The parameters established here will aid the realisation of waveguides and DMLs for $Ca^+$ wavelengths. Our development of the 3D FLDW waveguides for blue wavelengths was performed at 405 nm (see supplementary material), so we expect a straightforward extension to 397 nm. $Yb^+$ is another species used in quantum information research but is more challenging in this context because cooling light is at 369 nm. Extensions of the techniques presented here to the ultraviolet will require further study of suitable materials for high transmission [41, 42]. Similarly, the module presented here covers all the wavelengths used to address $Ba^+$, with the exception of the 1762 nm quadrupole transition [43]. For such a long wavelength, the challenge is to create DML profiles deep enough to impart a full $2\pi$ phase shift on the wavefront to achieve high diffraction efficiencies.

*4.2 Performance improvements*

To demonstrate the capability of the developed fabrication process, DML parameters were chosen which yielded small diameter beams suitable for addressing small numbers of ions at each trapping site. For interactions of lasers with trapped ions, larger beam diameters are often used to allow for more uniform illumination of longer ion strings. Larger beam sizes would require DMLs with less stringent fabrication tolerances than those achieved here. The observed systematic offset error of the NIR beams in the intersection plane is likely to originate from a 1% error in the beams' *k*-vectors. These small errors are believed to arise from the DML model and fabrication development being optimised for blue wavelengths, where excellent overlap was demonstrated and the lithography was more challenging. As the errors are consistent between beam sets, we expect them to be correctable with a similar optimisation of the DML model at NIR wavelengths. Despite the slight measured offsets, the resulting intensities of the NIR beams at the ion are sufficient for the driving of their target atomic transitions.

The present limit to DML diffraction efficiency is non-optimised surface-relief depths, which could be improved with further process optimisation. In a separate prototype DML array (not used in the assembly), diffraction efficiencies of ~67% (~1.7 dB) were measured for 422 nm lenses. For NIR wavelengths, process development is required to increase the etch depth and hence the diffraction efficiencies. Parameters to investigate are the etch selectivity of PMMA over $TiO_2$ (3.9), as well as increased resist thickness (2.5 µm). Increasing the latter will reduce the fabrication resolution of the NIR DMLs, however these structures have a lower aspect ratio (0.2) than for the blue wavelength DMLs (0.25) shown here; thus, we anticipate NIR diffraction efficiencies of > 67% could be achieved. Etch selectivity improvements by further optimisation of the chemical and physical etching would not compromise fabrication resolution.

The effect of crosstalk intensity between a site illuminated for state detection and a separate, unilluminated trapping site is considered. In assuming that one site is illuminated with an intensity suited to efficient Doppler cooling, then a relative intensity $I_{cross}/I_0 = 1 \times 10^{-3}$ corresponds to a photon scattering rate of ~2.5 ms$^{-1}$ for an ion stored in an unilluminated site. Future developments will need to increase the diffraction efficiency of the DMLs to improve this performance, since the largest contribution to observed crosstalk illumination is undiffracted light. This will be essential to minimise crosstalk-induced decoherence in the ion's quantum state during a protocol. For some protocols, it may be necessary to selectively quench qubit states in one zone (with 1033 nm), while preserving qubit coherence in other zones. Characterisation of such protocol-specific crosstalk errors will be required in future work using trapped ions as the probe.

Incorporating all necessary beams in a future module will require further developments on polarisation performance. To realise circular or linear polarisation with a microtrap in practice, in-line compensation will be required at the input of those channels for which is it most critical, For example, qubit control beams may require linear polarisation at a level not achievable with the present module. Optical pumping beams will require linear or circular polarization depending on the ion species. Optimised compensation would be achieved with the input VGA fibre at a constant position and temperature, using trapped ions as a discriminant. This approach will limit scalability, so more speculative building blocks to enhance the system could be considered. Recent fabrication developments include stress members written on either side of the waveguide in fused silica to induce a birefringence of the same magnitude as in PM fibre [44], and waveplates written into waveguide circuits for quantum photonics [45].

*4.3 Proximity of module to microtrap*

For its intended application, the micro-optical assembly will be in proximity to an ion microtrap. Dielectrics can accumulate charge and so distort the trapping potential by a significant amount if close enough to the trap. We have considered this by constructing a finite-element model (in *COMSOL Multiphysics*®) of the trapping pseudopotential arising from the microtrap device. We found that a grounded conductive coating on the $TiO_2$ surface, with uncoated DML apertures (100 μm × 100 μm) will permit a >1 mm working distance of the micro-optical assembly from the microtrap, without displacement of the ion position from the RF null (at the 210 nm resolution of the simulation mesh). Displacements due to working distances of 400 μm can be corrected with potentials of a few volts applied to the trap compensation electrodes. In other work [46], DMLs (140 μm × 140 μm) for fluorescence imaging were assembled in a surface electrode microtrap with the metal-masked lens substrate 165 μm from the ions; stray fields were readily compensated. We speculate that the 2 mm working distance of the present module design (i.e. > 8 times the ion-electrode distance) is a conservative value in this context.

Increasing the module working distance beyond 2 mm would require larger area DMLs with a greater depth-to-pitch aspect ratio for efficient performance. The increased DML area will limit scalability to large numbers of lenses in an array. This suggests that a practical upper limit to working distance is not much more than 2 mm. This limit would likely prevent the module's use in delivering laser beams across the surface of a microtrap with a 2D electrode geometry. In contrast, it could be applicable to a surface trap device with a clear front-to-back aperture, such as the Sandia HOA microtrap platform [47].

### *4.4 Towards an optically interconnected 3D microtrap*

Assembling the micro-optical module requires precise alignment and bonding of the parts. The module also needs to be accurately spaced from, and aligned and bonded to, the microtrap chip to form a hybrid system. These processes will require the use of photonics packaging tools and optical diagnostics to ensure accuracy. The discriminant for aligning the VGA to the waveguide chip is optical transmission of two waveguides (one at each extreme of the array). Beam overlap is the discriminant for aligning the DML array to the waveguide chip. We find that a 1 μm offset in DML position results in a ∼ 1 μm offset of the beam from its intended position, and that overlapping three beams results in the remainder of the array in the correct position. For a large-scale array, overlapping beam sets at extremes of the array would ensure alignment of the entire array. Knowing the DML substrate thickness to ± 5 μm enables the DMLs to be designed so that the substrate can be bonded directly to the waveguide chip and achieve the desired beam specification.

A micromachined spacer is needed to realise the required working distance ($z$ direction) of the module from a microtrap chip, and to facilitate bonding the two together. The spacer material thickness (e.g. 1800 μm) must be known with a tolerance of ± 5 μm to design the DML array and achieve beam overlap at the desired position. Overlapping the modes of auxiliary optical channels in the module onto markers on the surface of the microtrap chip will ensure lateral ($x,y$) alignment; 3 auxiliaries will be sufficient to align all outputs on their respective trap segments. The accuracy of beam overlap and positioning demonstrated here is adequate for this purpose. A more sophisticated approach to alignment could follow precision passive mechanical alignment principles [48]. Related work has shown sub-micron repeatability and accuracy in assembling a micro-optical bench [49], and the use of micro-mechanical stops for self-alignment of photonic dies [50]. Self-alignment principles feature in a proposal for mounting an optical fibre cavity onto a surface electrode ion microtrap [51]. Developments to test alignment principles are in progress.

Technical challenges around containment in UHV are two-fold. Development of vacuum fibre feedthroughs for many optical channels will be required, as is the case for alternative approaches using waveguides embedded in microtraps with 2D surface electrode geometries. Bonding components using UHV compatible epoxies is another aspect; from other work we find that that the most promising approach is to bond the components with high viscosity epoxy away from the optical paths so that it does not impact transmission.

*4.5 Scalability considerations and future directions*

With the different components aligned and bonded to one another, and subsequently to a microtrap chip, many beams will be aligned to the device using a single set of components. In contrast, bulk optical systems require a separate alignment procedure for each beam which is not scalable to a large number of microtrap zones. In our approach, the alignment complexity remains largely the same irrespective of number of beams so is inherently more scalable than bulk systems. A possible limit in scaling the module to illuminate a large number of trapping zones is the area to accommodate the required array of DMLs. Our design calculations suggest that it is feasible to illuminate each alternate electrode zone in a microtrap array with three beams for cooling and one each for optical pumping, repumping, quenching and qubit transitions. Given that the unilluminated zones will perform as axial barriers and transfer zones, then scaling to illuminate large arrays without a limit to axial extent seems feasible. Should minimised insertion loss be desired (i.e. correcting the VGA fibre core position deviation) when scaling to hundreds of fibres over many devices, a greater degree of automation will be required for the fabrication process.

In future the principles of this work could be applied to fluorescence imaging of trapped ions, which is the basis for quantum state detection. Fluorescence detection of individual ions has been achieved using components such as a binary microfabricated Fresnel lens [52] and a multimode optical fibre [53]. A micro-optical system operating in reverse to that presented here could detect fluorescence from multiple trapping zones, with one DML and waveguide channel per zone. However to achieve sufficient signal-to-noise, the realisation of efficient DMLs with large NA (~0.2) will be technically challenging. Once overcome, an assembly could be fabricated which provides precise, stable addressing and fluorescence collection in a single optical platform allowing for novel experiments with trapped atomic ions.

## 5. Conclusion

The micro-optical module demonstrated here exhibits operating characteristics that are appropriate for the optical addressing of ions stored in a segmented microtrap array. The principal enabling components of the module are the remapping laser-written waveguide chip and the 2D array of DMLs. Each output laser beam propagates precisely to a DML which then directs the beam with a unique *k*-vector to the target intersection point, where spatial overlap with other beams in the set is excellent. This performance, achieved with blue and NIR wavelengths, meets the initial optical design specification. The principle of the system demonstrated here is suited to the wavelengths required for a range of atomic species. The future aim will be to combine such a module with an ion microtrap, to form a hybrid system of components for atomic quantum technologies. We estimate that the module demonstrated here occupies a volume $\sim 10^4$ times smaller than the bulk opto-mechanical system we use for irradiating a pair of microtrap zones in our present apparatus [54].

## Funding and disclosures

*Funding*

M.L.D. was supported in this work by a UK DSTL Studentship and the EPSRC Centre for Doctoral Training in Quantum Engineering (EP/L015730/1). We acknowledge the support of the UK government department for Business, Energy and Industrial Strategy through the UK national quantum technologies programme and the UK EPSRC grant QuPIC (EP/N015126/1).

The work was performed in part at the OptoFab node of the Australian National Fabrication Facility utilizing Commonwealth as well as New South Wales state government funding. Funding through the Australian Research Council Discovery Program (DE160100714) and the Australian Research Council Centre of Excellence (CE110001018) is also acknowledged.

*Disclosures*

The authors declare no conflicts of interest.

# A micro-optical module for multi-wavelength addressing of trapped ions: supplementary material

**Matthew L. Day,**[1,2,3,†] **Kaushal Choonee,**[2] **Zachary Chaboyer,**[4] **Simon Gross,**[4] **Michael J. Withford,**[4] **Alastair G. Sinclair,**[2] **and Graham D. Marshall**[1]

[1] Quantum Engineering Technology Labs, University of Bristol, BS8 1TL, UK
[2] National Physical Laboratory, Teddington, TW11 0LW, UK
[3] Quantum Engineering Centre for Doctoral Training, University of Bristol, BS8 1TL, UK
[4] MQ Photonics Research Centre, Department of Physics and Astronomy, Macquarie University, New South Wales 2109, Australia
Email: alastair.sinclair@npl.co.uk

† Present address: Institute for Quantum Computing, University of Waterloo, N2L 3G1, Canada

**This document provides supplementary information to "A micro-optical module for multi-wavelength addressing of trapped ions"** https://doi.org/10.1088/2058-9565/abdf38

## S1. Waveguide mode optimization

During development of the waveguide chip, lasers with the exact application wavelengths were not available at the fabrication facility. Thus fabrication parameters were chosen for maximal mode overlap of $\lambda = 405$ nm and $\lambda = 976$ nm light with the single-mode fibres in the array (S405-XP and 1060-XP respectively).
For each waveguide in the chip, the output mode was imaged and fitted by the Gaussian function

$$I(x,y) = I_0 exp(-x^2/2w_{0x}^2)exp(-y^2/2w_{0y}^2). \quad (S1)$$

The mode-field diameter for the input fibre was determined by averaging the measured mode-field diameter in several representative fibres. With the approximation that the input mode was a perfect Gaussian of that diameter, the overlap with the waveguide mode was calculated using the mode overlap integral

$$S = \frac{\iint \sqrt{I_1(x,y)I_2(x,y)}dxdy}{\sqrt{\iint I_1(x,y)dxdy \iint I_2(x,y)dxdy}}. \quad (S2)$$

Example data used to calculate the mode overlaps in the main manuscript are shown in Figure S1.

## S2. Beam profile and crosstalk measurements

An exposed board CCD was used to measure the profile and position of a freely propagating beam emerging from the DML array. With the CCD located at the intersection plane, the DML outputs were imaged sequentially to align the DML array to the waveguide chip. At a fixed waveguide-to-DML alignment, beam profiles at the two intersection points were recorded. This setup was also used to measure high dynamic range beam profiles via a method described previously [1, 2]; these profiles enabled bounds on crosstalk to be quantified. The crosstalk intensity is determined from the average of the 10 pixels (spanning 11.2 µm) centred on the unilluminated intersection point, i.e. either $\Delta x = +350$ µm or $-350$ µm from intersection point A or B respectively. The maximum background intensity is taken from all measured pixel values along the $y = 0$ axis, which provides a more complete picture on worst-case crosstalk values ($6.7 \times 10^{-3}$) elsewhere in the microtrap. Large-area 2D HDR images of the lens outputs are shown in Figure S2.

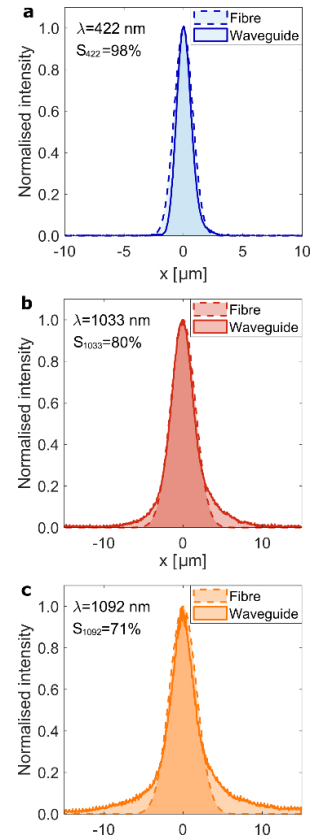

Fig S1. Measured beam profiles of waveguide and fibre modes at (a) 422 nm, (b) 1033 nm, (c) 1092 nm.

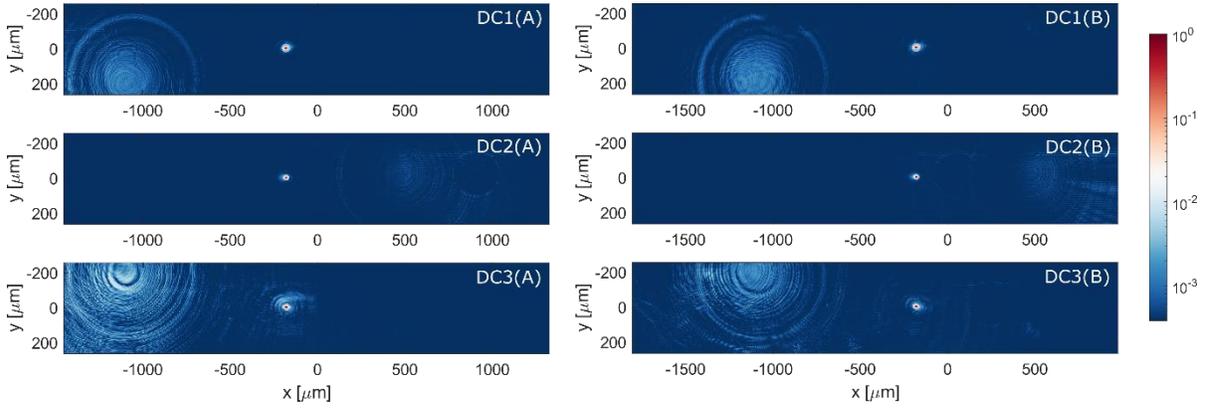

Fig S2. Large-area HDR images of the lens outputs, showing the locations of scattered laser light relative to the intersection points. In all cases the largest contribution to the background scatter is from the undiffracted light from the diffractive microlenses.

The accuracy of the HDR stitching process was ensured by calibrating the exposed board CCD over 4 magnitudes of intensity. The calibration process involved recording the camera response, $R$, as a function of laser power, $P$, for four different exposure times, $T$. The measured response at each exposure was fit with the function

$$R(P,T) = \left[ K(P - P_0)T + (R_0(T))^{\frac{1}{\gamma}} \right]^\gamma \quad (S3)$$

where $K$ is a gain, $P_0$ is the optical power of the incident beam above which a signal is detectable by the sensor, $R_0$ is the response of the sensor under no illumination and $\gamma$ is the gamma-correction factor. The values of $K$ and $P_0$ were left as free fit parameters, while $R_0$ was measured for each exposure time and $\gamma = 0.29$. (Initially, $\gamma$ was left as a third free parameter in fitting each dataset, after which the average value of $\gamma = 0.29$ was used in the two-parameter fits.) After fitting, a linear mapping between camera pixel value and relative intensity is calculated from the inverse of Eq. (S3) and then normalised to the peak power to provide relative intensity as a function of sensor response. The linearised camera signal response for all exposure times is shown in Figure S3. The linearisation procedure was applied to the stitched beam profiles to form accurate HDR beam profiles.

### S3. Waveguide position correction

Waveguide arrays were fabricated with and without applying $(x,y)$ positions corrections to compensate for fibre core deviations. Power efficiency measurements of two arrays fabricated in the same chip are presented in Figure S4, demonstrating improvements in coupling efficiencies for all waveguides, with improvements ranging from increases of 10 to 30 %.

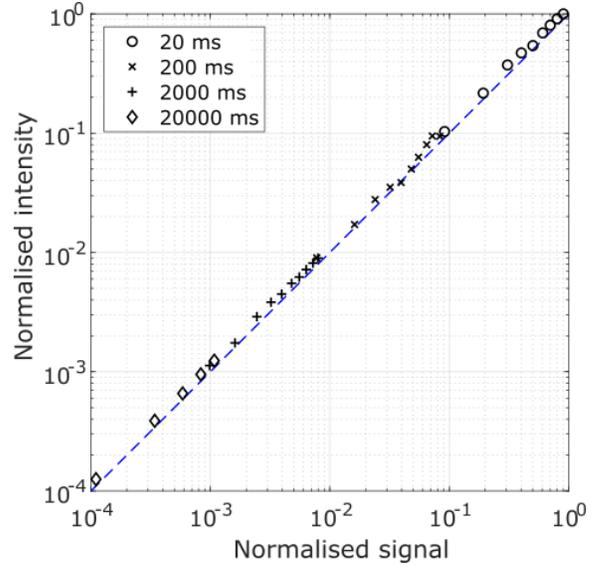

Fig S3. Signal response of an exposed board CCD camera after linearisation. It can be seen that linear response curves can be measured over 4 orders of magnitude.

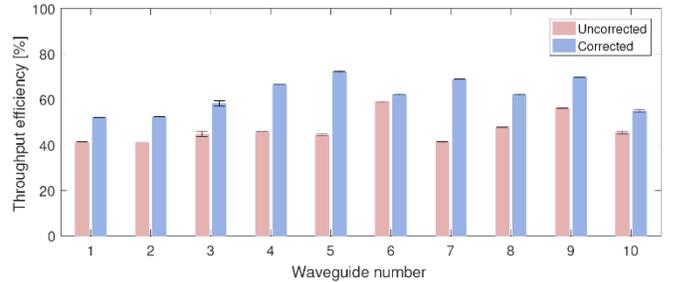

Fig S4. Measured throughput power efficiencies of two waveguide arrays fabricated in the same glass chip using the same fibre array. One waveguide array was fabricated without correction for known fibre core position offsets, one waveguide array was corrected for these offsets. In all waveguides, the correction improves power throughput.